\documentclass[iop]{emulateapj}
\usepackage{url}
\usepackage{epsfig}
\def\arcsec{$^{\prime\prime}$}
\newcommand\kms{km s$^{-1}$}

\newcommand\ha{$\rm{H}\alpha$}

\newcommand\tten{$\times$10$^{12}$}
\def\jcap{{JCAP}}

\shorttitle{NGC 4258 Satellites}
\shortauthors{Spencer et al.}

\begin{document}

\title{A Survey of Satellite Galaxies around NGC 4258}

\author{Meghin Spencer\altaffilmark{1}$^,$\altaffilmark{2}, 
         Sarah Loebman\altaffilmark{1}$^,$\altaffilmark{3},
        Peter Yoachim\altaffilmark{4}} 

\altaffiltext{1}{Department of Astronomy, University of Michigan, 500 Church Street, Ann Arbor, MI 48109}
\altaffiltext{2}{Correspondence should be addressed to meghins@umich.edu}
\altaffiltext{3}{Michigan Society of Fellows}
\altaffiltext{4}{Department of Astronomy, University of Washington, Box 351580, Seattle, WA 98195}

\begin{abstract}
We conduct a survey of satellite galaxies around the nearby spiral NGC
4258  by combining  spectroscopic observations  from the  Apache Point
Observatory 3.5-meter telescope with SDSS spectra. New spectroscopy is
obtained for 15 galaxies. Of  the 47 observed objects, we categorize 8
of them  as probable satellites, 8  as possible satellites,  and 17 as
unlikely to  be satellites. We do  not speculate on  the membership of
the  remaining 14  galaxies due  to a  lack of  velocity  and distance
information.  Radially integrating our  best fit  NFW profile  for NGC
4258 yields  a total  mass of 1.8$\times$10$^{12}$  M$_{\odot}$ within
200  kpc. We  find that  the  angular distribution  of the  satellites
appears to be random, and  not preferentially aligned with the disk of
NGC 4258.  In addition, many  of the probable satellite  galaxies have
blue $u-r$  colors and  appear to be  star-forming irregulars  in SDSS
images; this stands  in contrast to the low  number of blue satellites
in the Milky Way and M31 systems at comparable distances. 
\end{abstract}

\keywords{galaxies: kinematics and dynamics --- galaxies: individual (NGC 4258) --- galaxies: structure --- galaxies: satellites}

\section{Introduction}

The  number,  type  and  spatial  distribution  of  satellite  systems
surrounding host galaxies  can serve as a unique  probe to theories of
galaxy formation.   In a $\Lambda$CDM universe, host  galaxies such as
the  Milky  Way (MW)  are  predicted to  reside  in  vast dark  matter
halos. In this paradigm, the  main galaxy is expected to be surrounded
by numerous satellite dark matter halos  that lie within a few 100 kpc
of the central  galaxy. It is reasonable to expect  that many of these
halos  host  luminous  dwarf  galaxies.  Such  satellite  systems  are
potentially   ideal   dynamical  tracers   of   the  underlying   mass
distribution  and  can  also   probe  the  effect  of  environment  on
morphology,  star formation, and  quenching.  Despite  the theoretical
utility  of a  well-sampled satellite  system, most  observations find
only a handful of satellites  per massive host outside the Local Group
(LG).
 
A limiting factor in  establishing robust samples of satellites beyond
the  LG is  the inherent  faintness of  these objects.   To circumvent
this, some  have taken a statistical approach,  stacking large numbers
of    similar   systems    and    inspecting   aggregate    properties
\citep{prada2003,zaritsky1993}.   However,  stacking  in  this  manner
makes it  impossible to study  variation between systems,  and thereby
restricts  the inferences  that  can be  drawn.   For example,  recent
observations of the MW and M31 satellite systems have found them to be
preferentially         aligned        in         extended        disks
\citep{kroupa2005,Conn2013,Ibata2013}.  Such results are impossible to
replicate  in  studies  that  stack  galaxies  and  requires  detailed
knowledge  of   particular  isolated  hosts  with   large  numbers  of
satellites to verify the global significance.

In this light, one recent study \citep[hereafter K11]{kim2011} focuses
on the MW-like, barred spiral galaxy  NGC 4258 (also known as M106) to
identify a large sample of satellite galaxies.  K11 use MegaCam on the
Canada France  Hawaii Telescope to observe  a 1.7 by  2.0 degree field
around NGC  4258 (roughly out  to a projected  radius of 130  kpc) and
find  16  candidate  satellite   galaxies  and  5  probable  candidate
satellite  galaxies.  These were  selected based  on the  existence of
literature  radial velocities, resolvable  stars, and/or  extended and
faint surface brightness structure.  This catalog of 21 galaxies spans
a wide range of morphologies, from dSph, dE, Sd, to Irr.  Most of them
have surface brightness  profiles that are fit well  by an exponential
and have negligible color gradients.  Additionally NGC 4258 satellites
follow  the Schechter luminosity  function with  a faint-end  slope of
-1.19$^{+0.03}_{-0.06}$,   which  is   steeper  than   the   LG  slope
(-1.06$\pm$0.03)     but     flatter     than    the     M81     slope
(-1.29$^{+0.07}_{-0.03}$).

We revisit this work by spectroscopically observing a subset of the 16
satellite  galaxies identified  in  the K11  sample  to determine  the
prevalence of  foreground/background galaxy contamination.   We expand
the K11  survey to include  additional galaxies detected in  the Sloan
Digital Sky Survey (SDSS) that are also spatially near NGC 4258.

NGC  4258 is  similar  to the  Milky Way  in  that it  has a  terminal
rotational  velocity of  208 \kms\  \citep{erickson1999}, is  a barred
spiral galaxy,  and has an  average B-band surface brightness  of 23.1
\citep[][]{kara2013}.\footnote{http://www.sao.ru/lv/lvgdb/} Due to its
relatively   nearby  location   \citep[7.6~Mpc,][]{humphreys2013}  and
interesting  inner  disk  morphology  \citep[resolved AGN  and  warped
  accretion disk,][]{caproni2007,martin2008},  it has been  the source
of  a  wide  array  of  galactic  studies.   Significantly,  NGC  4258
possesses a water  maser, which tightly constrains the  distance to it
with  an  error  of  only  $\sim$$3\%$  \citep{humphreys2008}.   Other
distance measurements made  with Cepheid variables and tip  of the red
giant   branch    (TRGB)   magnitudes   have    reasonable   agreement
\citep{humphreys2013}.

Aside  from   assessing  the  significance   of  foreground/background
contamination in  a photometrically  derived satellite sample,  we see
three  primary motivations to  extend the  K11 study  of the  NGC 4258
satellite  system.   One,  using  a Jeans  equations  based  technique
developed by \citet{watkins2010}, we can draw constraints on the total
mass  of NGC  4258.  Two,  we can  assess if  there is  a preferential
orientation for the satellites surrounding NGC 4258, as has been found
for  the MW \citep{kroupa2005},  M31 \citep{Conn2013,  Ibata2013}, and
M81 \citep{chiboucas2013} Three, it has long been known that the color
and morphology of MW and M31 satellites vary as a function of distance
from   respective   hosts   \citep[see,  for   example,][]{mateo2008}.
However, it has  yet to be established if  this apparent environmental
effect holds outside the LG.   Given its relative isolation from other
massive  perturbers  and   numerous  candidate  satellites,  NGC  4258
provides an ideal testbed of the significance of proximity to the host
galaxy to color and morphology of satellites.

For these reasons,  we build upon K11's work  to assemble and leverage
the  most complete list  of satellite  galaxies surrounding  a MW-like
system outside  the LG.   The structure of  this paper is  as follows:
Section 2 details our data collection/reduction process, including the
target   selection,  observations,   data  processing,   and  velocity
measurements.  Section  3 provides a  discussion of how  we categorize
the satellites  and calculate  the host mass.  In the same  section we
present a  discussion of the  angular distribution and color  range of
the satellites.  Section 4 contains concluding remarks. 

\section{Data Collection} \label{data collection}

\subsection{Target Selection} \label{target selection}
\begin{figure*}
\centering
\plotone{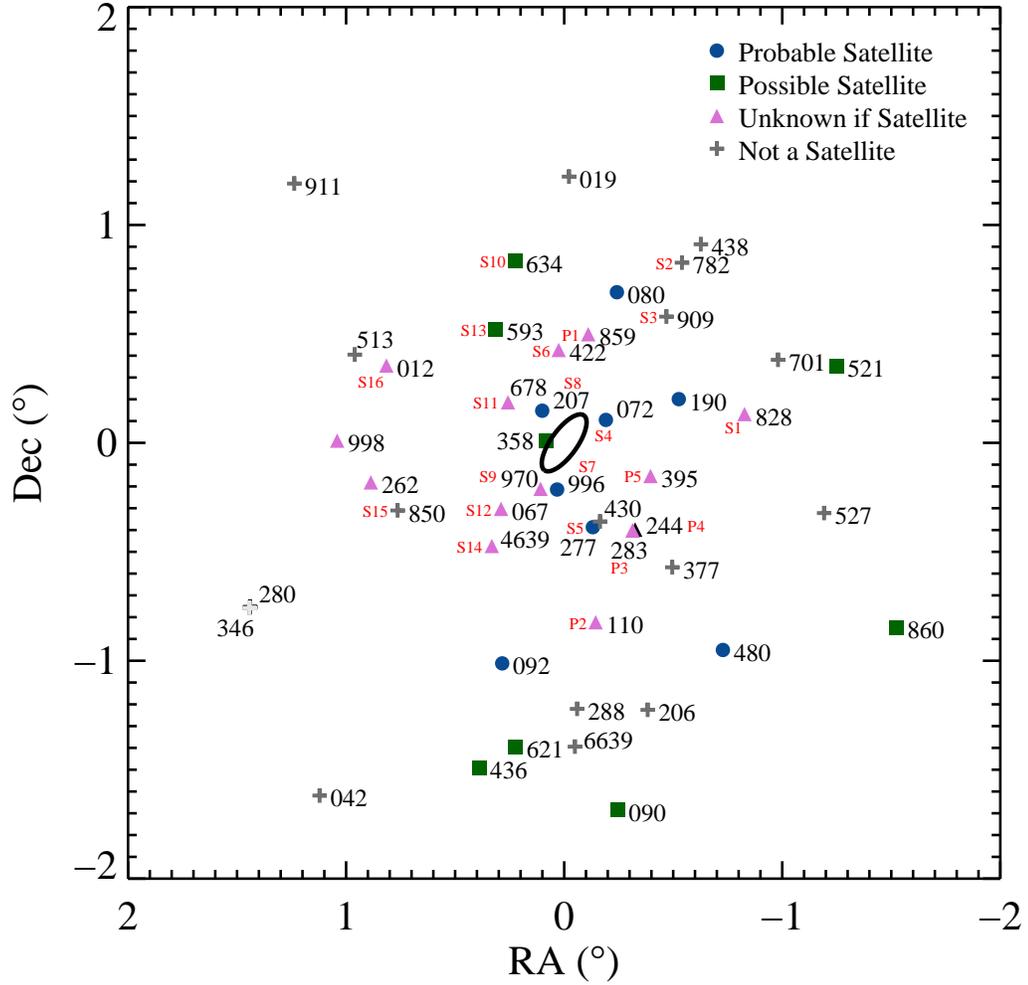}
\caption{Locations of the considered galaxies on the sky. North is up;
  east  is  left.  A  large  central  oval  represents  the  location,
  inclination, and size  of NGC 4258. Each galaxy  is labeled with the
  last three or four digits of  its SDSS ID that uniquely identify it,
  and the K11  ID when available.  Various plot  symbols represent the
  membership categorizations that we make in Section 3.1. Blue circles
  are  probable satellites  (classified  as ``Y");  green squares  are
  possible satellites (classified as ``M"); gray pluses are background
  galaxies (classified as ``N"); pink triangles are galaxies that lack
  velocity information (classified as ``X").  \label{name}}
\end{figure*}

We combine the K11 catalog  of 16 potential satellites with additional
candidates drawn from  the SDSS database to produce  a list of targets
for spectroscopic followup. To  select potential targets from SDSS, we
searched within 2 degrees (projected radius $\sim250$ kpc) of NGC 4258
and  required  that  the   r-band  Petrosian  radius  be  larger  than
5\arcsec\ and the  photometric redshift be less than  0.1. Minor color
cuts were applied to ensure  targets had galaxy-like colors. This list
was  sorted  by r-band  brightness,  and  then  targets were  manually
inspected to  remove any objects that  were artifacts of  SDSS such as
shredded galaxies  or bright star  halos. Satellites were  then ranked
based on  r-band magnitude, projected  distance of the  satellite away
from the host, and  photometric redshift, using the following formula:
score                                                                 =
$\sqrt{(\frac{r}{r_{max}})^{2}+(\frac{d}{d_{max}})^{2}+(\frac{z}{z_{max}})^{2}}$. The
30 highest  scoring galaxies as well  as the K11  satellites served as
the basis for  our observations. Due to limited  observing time we did
not obtain spectra for all of these objects. See Figure~\ref{name} for
galaxy distribution in right ascension and declination relative to NGC
4258   and    Table   \ref{data}   for    relevant   properties   from
SDSS\footnote{http://sdss3.org}  and NASA/IPAC  Extragalactic Database
(NED)\footnote{http://ned.ipac.caltech.edu/}.

\subsection{Observations and Data Reduction}

We used  five half-nights  of observing time  on the  3.5-meter Apache
Point Observatory  (APO) telescope with the  Dual Imaging Spectrograph
(DIS)  to observe  29 targets.  The  wavelength was  centered on  6799
\AA\ for the red channel and 4502 \AA\ for the blue, with each channel
covering  about  $\sim$1180 \AA.  The  high  resolution 1200  lines/mm
grating was used  with a slit width of 2''.  Observations were made in
February, April, and  May of 2011. Exposure times ranged  from 5 to 20
minutes.

Ten bias frames were averaged  and subtracted from all images. Nightly
quartz lamp  dome flats were normalized  by a 9th, 10th  or 11th order
polynomial  and   divided  out  of   the  science  images.   Arc  lamp
Helium-Neon-Argon spectra were taken after  slewing to a new target to
correct for any  small distortions in the mirror  and instrument while
the telescope was  in motion. Arc lamp frames  were used to wavelength
calibrate   the   corresponding   science   spectra.   Standard   star
observations taken at  the beginning and/or end of  each night.  These
flux  standards  were  used   to  eliminate  the  instrument  response
signature in the spectra.  Typical  seeing was 1.6", and the standards
were observed at low airmass.   The background was subtracted by using
regions off-target but along the slit. 

\begin{figure*}
\centering
\plotone{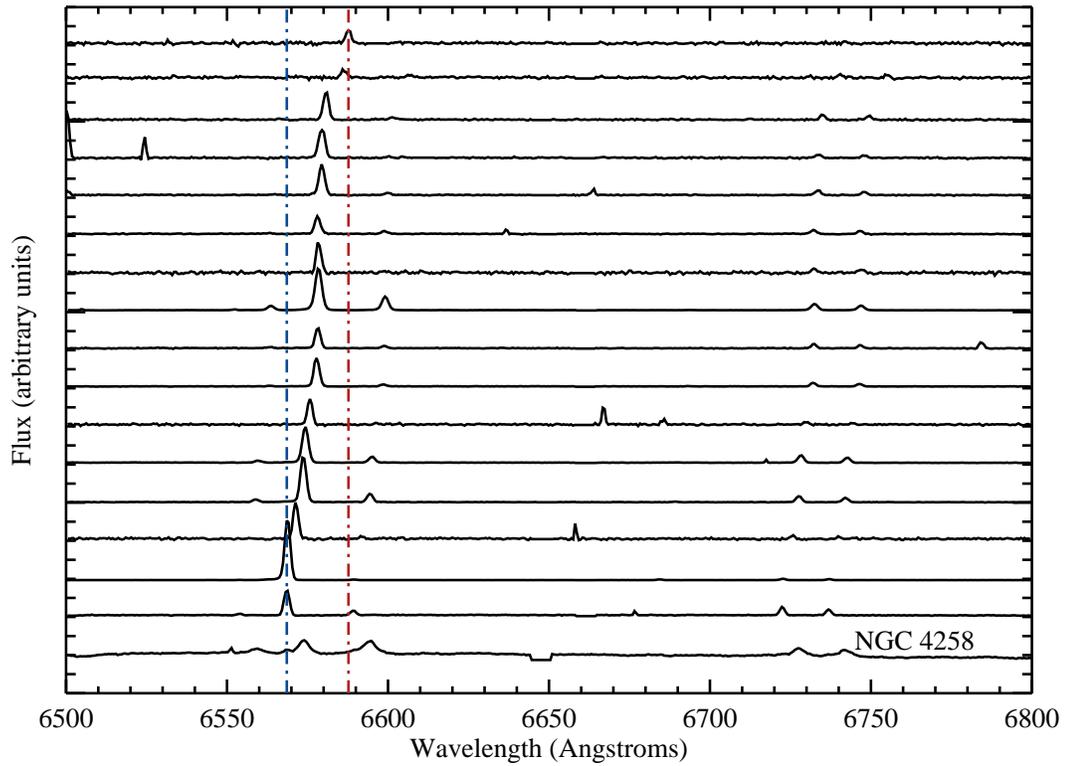}
\caption{Spectra for  16 of the observed galaxies  that have redshifts
  similar to that of NGC 4258.  Spectra shown are from the red channel
  of  DIS.  The   spectrum  of  the  host  galaxy   is  shown  at  the
  bottom.  H$\alpha$ is the  large emission  line on  the left  and is
  bracketed  by  the  [NII]  doublet.   On  the  right  is  the  [SII]
  doublet. These five emission  lines were used to determine redshifts
  of  the  galaxies. Vertical  red  and  blue  dashed lines  mark  the
  location  of  H$\alpha$  for   the  highest  and  lowest  redshifted
  satellites respectively. \label{specs}}
\end{figure*}

Spectra  were  corrected  to  the  heliocentric rest  frame  and  were
combined when repeat observations were made on the same night. Spectra
from  the   red  and  blue   channels  were  reduced  with   the  same
method. Completely reduced spectra have  a scale of 0.58 \AA/pixel for
the red channel and span a wavelength range of $\sim$6210 to 7390 \AA;
the scale for  the blue channel is .61 \AA/pixel,  spanning a range of
$\sim$3870 to 5130 \AA. See Figure \ref{specs} for a sample of reduced
spectra.

Objects observed  with APO are  listed in Table \ref{data}  along with
the  J2000  right   ascension,  declination,  SDSS  r-band  magnitude,
measured     heliocentric    radial    velocity,     and    literature
distances. Galaxies  will hereafter be  referred to by last  three (or
four) digits of their SDSS DR7 object  ID. For a full SDSS ID refer to
Table \ref{data}.

\subsection{Velocity Measures}
Depending on the  size of the target galaxy, we  extract between 1 and
20 spectra along the  spatial dimension.  For each extracted spectrum,
we use Penalized Pixel-Fitting (pPXF) \citep{Cappellari04} and Gas AND
Absorption  Line Fitting (GANDALF)  \citep{Sarzi06} algorithms  to fit
stellar velocities  and emission line  velocities in both the  red and
blue  spectra.  For  stellar  templates,  we use  the  single  stellar
population galaxy  models from \citet{Vazdekis10}.   For galaxies that
were  spatially  extended, we  fit  rotation  curves  to the  measured
velocities to find  the dynamical centers of the  systems and systemic
velocities.

The  red  emission  lines  (\ha,  [SII])  are  usually  the  brightest
features. The  red spectra  are also better  calibrated since  we only
have  $\sim$5 arc-lines  in  the  blue.  We  therefore  adopt the  red
emission line  velocity, or if it  is not available,  the blue stellar
velocity as the systematic velocities of our galaxies.  

Since many  of these  galaxies are spatially  extended, we  expect the
primary  source of  error  to be  due  to imperfect  placement of  the
spectrograph slit.  Given the high signal  in most of  our spectra, we
expect velocity errors  of order 10 \kms.  Formally,  the errors could
be smaller, but we expect systematic errors (slit-placement, kinematic
vs  photometric center, etc.)  will limit  us as  well. We  assume the
maximum in  the stellar continuum corresponds to  the dynamical center
of each galaxy. 

Our velocities  agree with the  available SDSS values within  the root
mean square  scatter of 10  \kms\ in all  but three cases. Two  of the
galaxies (SDSS  IDs 911 and  621) have bright off-center  star forming
regions which  were targeted  by SDSS. The  last galaxy  (782) remains
discrepant;  despite being  a relatively  bright object,  none  of the
measurements (ours, SDSS, or NED) agree particularly well.

Our  velocity for  NGC 4258  also  agrees with  literature values.  To
derive this, we  carefully extracted the spectrum from  a small central
region, as NGC 4258 has a very sharply rising rotation curve which can
skew the result if the extraction is not symmetric.

\section{Analysis}
Below we present three findings  that were made with the combined SDSS
and APO data.  The first section describes how  we categorize galaxies
as probable  and possible  satellites versus background  galaxies. The
second section  explains our mass  calculation of NGC 4258.  The third
section  examines  the angular  distribution  of  satellites, and  the
fourth section discusses the color of our satellite galaxies.

\begin{figure*}
\centering
\plotone{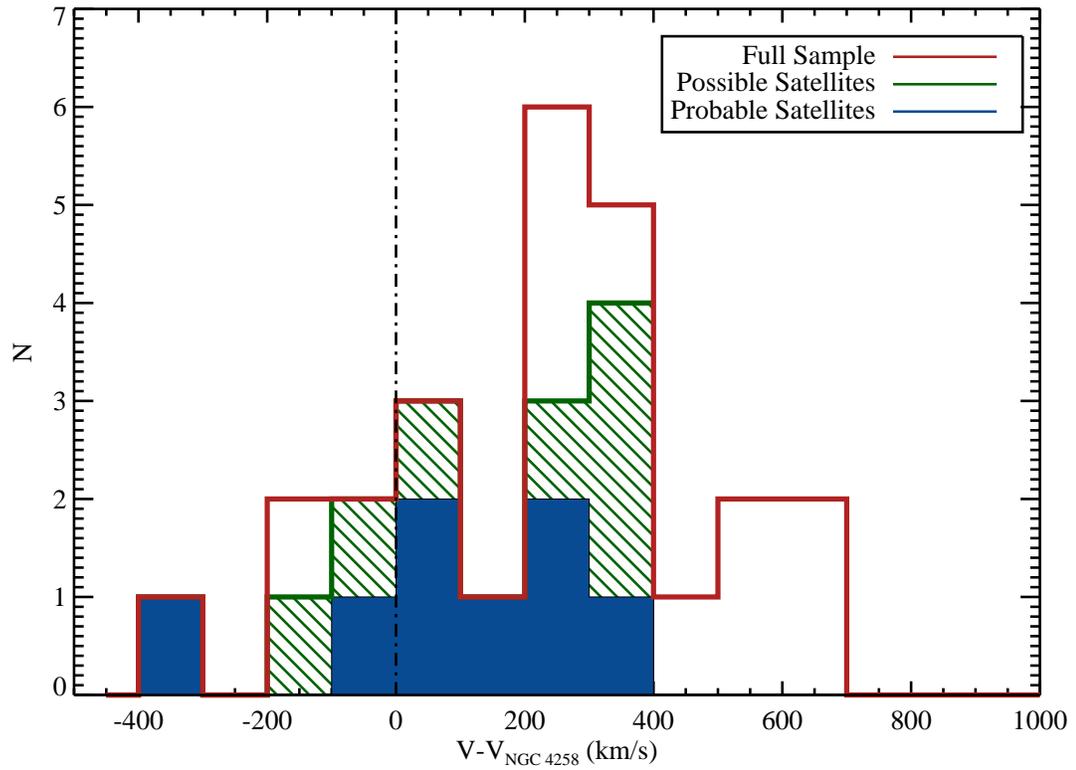}
\caption{Histogram   of  the   number  of   satellites  in   each  100
  \kms\  velocity bin.  The  vertical black  line  marks the  systemic
  velocity  of NGC 4258.  Notably, more  galaxies are  redshifted with
  respect  to NGC 4258  than blueshifted,  indicating the  presence of
  background galaxies \citep{zaritsky1992}.  In Section 3.1 we discuss
  how we  narrow down the  sample to probable satellites  (blue filled
  region) and  possible satellites  (green filled region).  The entire
  sample of galaxies is outlined in red. See Figure \ref{vvsr} for how
  these satellites are classified. \label{histogram}}
\end{figure*}

From  this point  forward, any  mention of  a velocity  refers  to the
velocity relative to NGC 4258. The relative velocity is defined as the
line of sight velocity of that galaxy minus the line of sight velocity
of NGC 4258. A histogram of  all these relative velocities is shown in
Figure \ref{histogram} as a red  line. It is immediately apparent from
this figure that there are many more galaxies that are redshifted with
respect to NGC 4258 than blueshifted. By symmetry, we expect a similar
number  of true  satellite galaxies  to be  redshifted  as blueshifted
\citep{zaritsky1992}. In  addition there is  a peak around  300 \kms\,
indicating that  there might be  some other structure just  beyond NGC
4258.  For  these  reasons,  we  conclude that  there  is  significant
contamination from  background galaxies in  our sample. Before  we can
perform any analysis on our sample  of galaxies or on the host itself,
we must discard non-members.

\subsection{Separating Satellites from Background Galaxies}

\begin{figure*}
\centering
\plotone{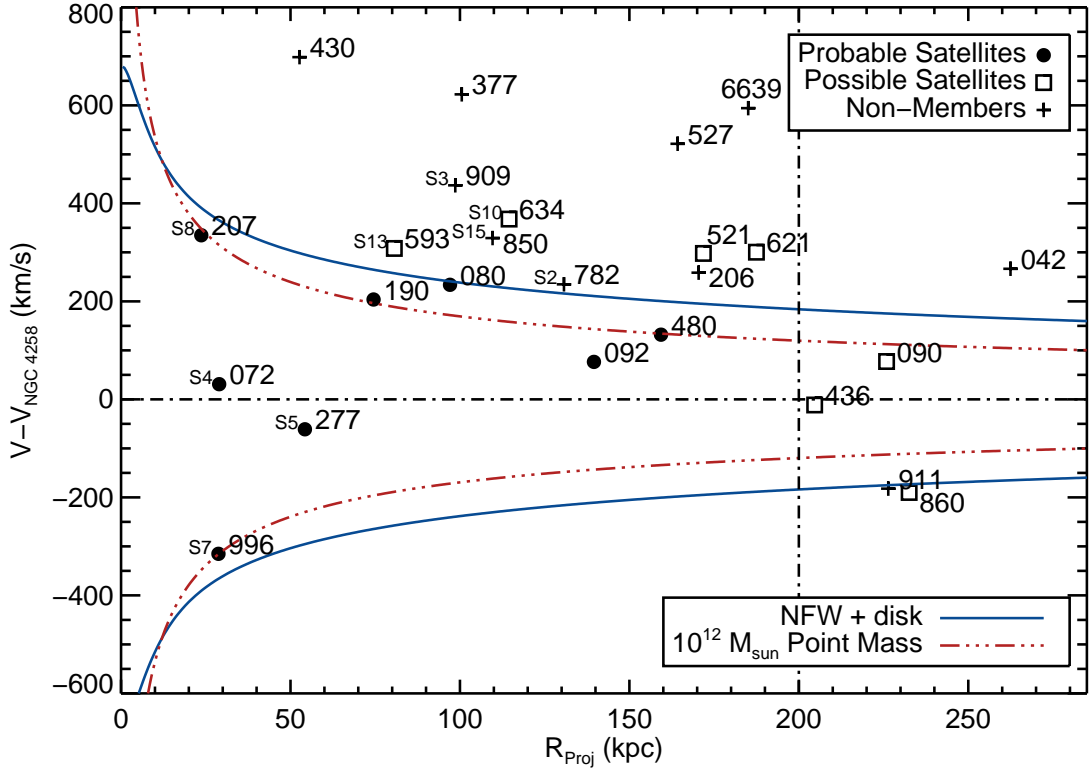}
\caption{Projected radius  vs line of  sight velocity with  respect to
  NGC 4258.  Potential satellite galaxies are marked  and labeled with
  the  last three or  four digits  of their  SDSS ID  and K11  ID when
  relevant.  Filled   circles  are  objects   classified  as  probable
  satellites;  open  squares   are  possible  satellites;  pluses  are
  non-members. Blue  and red  lines trace the  escape velocity  from a
  disk+NFW   profile    and   1$\times$10$^{12}$   M$_{\odot}$   point
  mass.  Escape velocity  profiles  are divided  by  $\sqrt{3}$ as  we
  assume  velocity isotropy  in  order to  compare  with the  observed
  galaxy  radial velocities.  Objects beyond  200 kpc  (vertical black
  dashed line)  are at  best listed as  possible rather  than probable
  satellites given their large projected radius \citep{barber2013}.}
\label{vvsr}
\end{figure*}

In order for a dwarf to be  considered a satellite it must be bound to
the host galaxy. That is, it  must have a total velocity less than the
escape velocity of the system  \textit{and} have a distance similar to
that of the host \citep{zaritsky1993}. To determine which of our dwarf
galaxies  could satisfy  the first  requirement, we  plot the  line of
sight velocity with  respect to the host against  the projected radius
(Figure \ref{vvsr}).

Next, we  calculate the  escape velocity as  a function of  radius for
three  different mass  distributions:  a point  mass,  an NFW  profile
\citep{navarro1996}, and  a Burkert profile  \citep{burkert1995}. Disk
components are included for the latter two profiles.

The NFW profile is described by  
\begin{equation}
\rho_{NFW}=\rho_{H}\frac{1}{x(1+x)^{2}},
\end{equation}
and the Burkert profile is described by
\begin{equation}
\rho_{Bur}=\rho_{H}\frac{1}{(1+x)(1+x^{2})}
\end{equation}
where $\rho_{H}$ is density scale, $x=R/R_{H}$ and $R_{H}$ is the core
radius for an NFW profile, which is of the order 10 kpc. We iterate on
$\rho_{H}$ and $R_{H}$ in our  analysis. The disk surface density that
we employ is
\begin{equation}
\sigma_{D}=\frac{M_{D}}{2\pi R^{2}_{D}}e^{-r/R_{D}}
\end{equation}
where the mass of  the disk, $M_{D}$, is 6$\times$10$^{10}$M$_{\odot}$
and the  scale radius of  the disk, $R_{D}$, is  2.6~kpc. Importantly,
since we only know 1-D velocities (radial velocities) of our galaxies,
we  assume   they  obey  velocity   isotropy\footnote{That  is,  $\sim
  V_{x}=V_{y}=V_{z}$;  it then  follows  that the  3-D  velocity of  a
  satellite is $|V_{tot}|=\sqrt{3V_{x}^{2}}$.}.  In this limit, we can
compare  our  line of  sight  velocities  with  the escape  velocities
predicted for each profile by dividing the profiles by $\sqrt{3}$.  We
have over  plotted $\pm(V_{esc}/\sqrt{3})$  for the three  profiles in
Figure~\ref{vvsr}.

Next, we  determine the  best fit parameters  to these profiles.  As a
first guess  we use  the parameters  for the Milky  Way as  defined by
\citet{nesti2013}.  Because the  NFW and  Burkert models  yield nearly
identical results, we opt to  use the NFW profile for further analysis
rather than both models. To  further refine the density models we need
to  know additional  information  about  the system.  If  there was  a
selection of  galaxies with reliable distance  measurements that match
NGC  4258  and have  velocities  within  a  reasonable range,  we  can
conclude that such  galaxies must be bound to  the system and leverage
them to  constrain the halo parameters.   This is the case  for two of
our 47 galaxies (996 and 207),  which have tip of the red giant branch
(TRGB)  distances  within the  errors  of  the  distance to  NGC  4258
\citep{munshi2007,kara2013}. More  importantly, they are  located near
the edge of the NFW escape  velocity profile and can be used to create
lower  limit boundaries for  the profile.  We increase  $\rho_{H}$ and
$R_{H}$  so that  these  two  satellites fall  within  the NFW  escape
velocity profile. Thus we assume density and radius values for the NFW
model  of  $\rho_{H}$=1.4$\times$10$^{7}  $ M$_{\odot}$/kpc$^{3}$  and
$R_{H}$=16~kpc.

We now use  our best-fit lower limit NFW  profile to determine further
satellite membership.  Any galaxy  that has a  line of  sight velocity
that  falls   within  the   NFW  $\pm  (V_{esc}/\sqrt{3})$   lines  in
Figure~\ref{vvsr} is  a strong candidate for being  a satellite galaxy
\citep[but   see][for   possible    confusion   at   large   projected
  radii]{barber2013}.   Eight   galaxies    fall   within   the   $\pm
(V_{esc}/\sqrt{3})$ boundaries  and have projected radii  of less than
$200$  kpc; they  are deemed  the  most probable  satellites, and  are
tagged with a  ``Y" for Yes in our catalog.  Galaxies that fall within
the $\pm  (V_{esc}/\sqrt{3})$ boundaries  but have projected  radii of
greater than $200$ kpc, are tagged with a ``M" for Maybe.

Next  we relax  our  initial assumption  about  velocity isotropy  and
consider the case when most of the total velocity is along the line of
sight\footnote{In  this case,  $|V_{tot}|  \sim V_{x}$};  a system  is
bound in  this scenario if the  line of sight velocity  is simply less
than the  total escape velocity. This  is a possible  but not probable
scenario. Hence, we categorize galaxies  with a line of sight velocity
greater than  $\pm (V_{esc}/\sqrt{3})$ but less than  $\pm V_{esc}$ as
possible satellites, and tag them with an ``M" for Maybe as well.

17  galaxies  fall  beyond  the  total  escape  velocity  profile.  We
categorize them  as non-members, and tag  them in our  catalog with an
``N" for No. The remaining  14 galaxies lack velocity information.  At
this  time,  we  draw  no  conclusion about  their  membership.  These
galaxies  are   indicated  in  Table  \ref{data}  with   an  ``X"  for
eXcluded. They are  not included in our subsequent  analysis but would
be strong candidates for further follow-up.

We consider the steps outlined  above a good first pass at determining
membership.  However, distance estimates  to these objects can help us
refine  our  individual   classifications.   The  literature  contains
distance measurements for  13 of our 47 galaxies. Here  we amend a few
of our previous Y/M/N/X categorizations  on the basis of these values,
as outlined below. 

Recall NGC 4258  is 7.6 Mpc away. Galaxy  850 was initially classified
as ``M;" however,  its average distance in NED is  15.7 Mpc from Tully
Fisher  and surface  brightness fluctuation  measurements; due  to the
large  separation in  distance,  it  is reclassified  as  a ``N''  for
background  galaxy.  Galaxies  909, 042,  and 206  have  average Tully
Fisher distances  in NED  of 17.9, 14.4,  and 21.05  Mpc respectively;
again, they are reclassified as ``N."  Another galaxy we reclassify is
782, which has  a single Tully Fisher distance in NED  of 21.2 Mpc; we
demoted from ``Y" to ``N."  Finally, galaxy 911 has a TRGB distance of
5.63 Mpc in NED. It was originally a ``Y" but is demoted to ``N" as it
is  a foreground  galaxy.  Any  galaxies unmentioned  retain  the same
membership classifications as  before. These final classifications are
listed in Table~\ref{data}.

We  add one  additional  galaxy to  our  list from  archival HST  data
analyzed by  L. Macri  and F. Munshi  \citep[see,][]{munshi2007}. They
present the TRGB  distance to galaxy 358. We categorize  it as an ``M"
galaxy since the  distance is similar to the  host galaxy, but refrain
from a higher  ranking because we do not  have a velocity measurement.
Additionally, Macri  and Mushi  also provide a  TRGB distance  for 207
that is consistent with NGC 4258.

It should be noted that seven of the galaxies we considered (067, 207,
4639, 422,  970, 678,  012) are  listed with distances  of 7.8  Mpc in
\citet{kara2013} \textit{based  upon K11's photometric  work}. 7.8 Mpc
is  the  distance to  NGC  4258  found  from Cepheid  measurements  of
\citet{newman2001}. We do not list these distances in Table \ref{data}
because, as we demonstrate  here, membership based on photometry alone
is uncertain.

When taking into consideration the available velocities and distances,
we  conclude that  4 of  the objects  classified by  K11  are probable
satellites (072, 207, 277, 996), 2 are possible satellites (593, 634),
3 are not satellites (782, 850,  909), and 7 are still uncertain (012,
067, 422,  678, 828, 970, 4639). We  note that 072, also  known as NGC
4248,    has    long    been    considered    a    satellite    galaxy
\citep{vanalbada1977}. See Table \ref{data} for mapping between K11 ID
and SDSS ID, as well as Figure \ref{name}.

We  conclude that  8  of the  galaxies  we consider  have the  highest
probability of being satellites. This represents 17\% of our sample. A
dwarf galaxy survey done by  \citet{carrasco2006} found only 78 out of
their 409 target galaxies  within four clusters were actually members,
equating to a 19\% yield. Since the surface density of galaxy clusters
is much larger than that of field galaxies, we consider the results of
the  \citet{carrasco2006} survey  as an  upper limit  for how  well we
expect to be  able to select system members.  However, we caution that
without  knowing both velocities  and distances,  it is  impossible to
state with absolute certainty whether  a galaxy is a satellite, and it
is still quite  possible that some of our most  probable sample are in
fact non-members.

To summarize, we adopt the following classification scheme:
\begin{itemize}
\item{Probable Satellites (``Y''): galaxies with distance measurements
  consistent  with NGC  4258  or  no known  distances;  line of  sight
  velocity  within $\pm (V_{esc}/\sqrt{3})$  and have  projected radii
  less than 200 kpc.}
\item{Possible Satellites (``M''): galaxies with distance measurements
  consistent  with NGC  4258  or  no known  distances;  line of  sight
  velocity greater than $\pm (V_{esc}/\sqrt{3})$ but smaller than $\pm
  V_{esc}$, or line of  sight velocity within $\pm (V_{esc}/\sqrt{3})$
  but with projected  radii greater than or equal to  200 kpc but less
  than 300 kpc.}
\item{Unknown   if   Satellites   (``X''):  galaxies   with   distance
  measurements  consistent  with  NGC  4258  or  no  known  distances;
  galaxies  with no known  line of  sight velocities;  projected radii
  less than 300 kpc.}
\item{Not  Satellites  (``N''):  galaxies with  distance  measurements
  inconsistent with NGC 4258 or velocities outside $\pm V_{esc}$.} 
\end{itemize}
Based  upon on  these criteria,  we categorize  8 objects  as probable
satellite  galaxies,  8 as  possible  satellite  galaxies,  and 17  as
non-members; for 14 galaxies we draw no conclusion.

\subsection{Halo Mass}

With this  collection of probable  satellite galaxies, we next  aim to
estimate the dynamical mass of NGC 4258. \citet{watkins2010} publishes
a set  of robust  mass estimators for  cases where only  the projected
radius and line of sight velocity of each target are known (as opposed
to true radii and peculiar  velocities). They assume the population of
satellites is spherically symmetric. The relevant equations are:
\begin{equation}
M=\frac{C}{G} <v_{\mathrm{los}}^2R^2> ,    C=\frac{(\alpha + \gamma - 2 \beta)}{I_{\alpha,\beta}} r^{1-\alpha}
\label{eq26}
\end{equation}

\begin{equation}
I_{\alpha , \beta}=\frac{\pi^{1/2} \Gamma (\frac{\alpha}{2} + 1)}{4 \Gamma (\frac{\alpha}{2} + \frac{5}{2})}[\alpha+3-\beta(\alpha+2)] .
\label{eq27}
\end{equation}

where M is the galaxy  mass, G is the gravitational constant, $\alpha$
is  a fiducial  radius at  which the  power-law approximation  for the
relative  potential  is  valid,   $\beta$  is  the  Binney  anisotropy
parameter  that   depends  on  the  tangential   and  radial  velocity
dispersions, $\gamma$  is the  power law index  of the  radial density
distribution  of  satellites,  r$_{out}$  is  the  upper  limit  of  a
gravitational field  that is scale-free, and $\Gamma(x)$  is the gamma
function, where  $\Gamma(x)=(x-1)!$.  We use $\alpha$  = 0 (satellites
move in a  large-scale mass distribution with a  flat rotation curve),
$\beta$  = 0  (isotropic  satellite  orbits), and  $\gamma$  = 2  (the
satellite density falls off as r$^{-2}$).

Incorporating  our  8  probable   satellite  galaxies  into  the  mass
calculation,    we   find    the   mass    of   the    host    to   be
3.1$\pm$0.7$\times$10$^{12}$  M$_{\odot}$  out  to  a  radius  of  200
kpc.  Including  the  8   probable  satellites  plus  the  3  possible
satellites that fall within $\pm (V_{esc}\sqrt{3})$ but have projected
radii  greater than  200  kpc, we  find the  mass  of the  host to  be
3.7$\pm$1.0$\times$10$^{12}$~M$_{\odot}$  out  to  240  kpc.   Current
estimates  for   the  MW  range  between   0.6\tten\  M$_{\odot}$  and
3.1\tten\      M$_{\odot}$      within      the     virial      radius
\citep{mcmillan2011,boylankolchin2013,barber2013}. 

An alternative way of calculating the mass is by integrating the NFW
density profile over an appropriate range of radius. Utilizing the NFW
profile derived in Section 3.1, we determine the total mass out to 200
kpc is 1.8$\times$10$^{12}$ M$_{\odot}$. Any uncertainty here is due
to the fact that we have 1-D velocities and projected radii.

\subsection{Satellite Distribution} \label{satellite distribution}
Several studies  of the  MW, M31, and  M81 have found  that satellites
exist     in     a      plane     centered     around     the     host
\citep{kroupa2005,Conn2013,Ibata2013,chiboucas2013}.   K11  find their
sample of 16  satellites are preferentially aligned along  the disk of
NGC 4258.  We  reassess these findings based on  our revised satellite
list.

A  rigorous evaluation  would consider  both a  satellite's  angle and
projected distance away from the disk before drawing conclusions about
the  angular dependence.   However, the  disk  of NGC  4258 is  highly
inclined relative to  our line of sight; this  minimizes the impact of
projection effects,  which could cause a satellite  elevated above the
disk to instead appear aligned with the disk.  Because of this we only
consider  the angular  separation  between a  satellite  and the  disk
plane. We use a position angle of 150$^{\circ}$ \citep{RC3}.

\begin{figure*}
\centering
\plotone{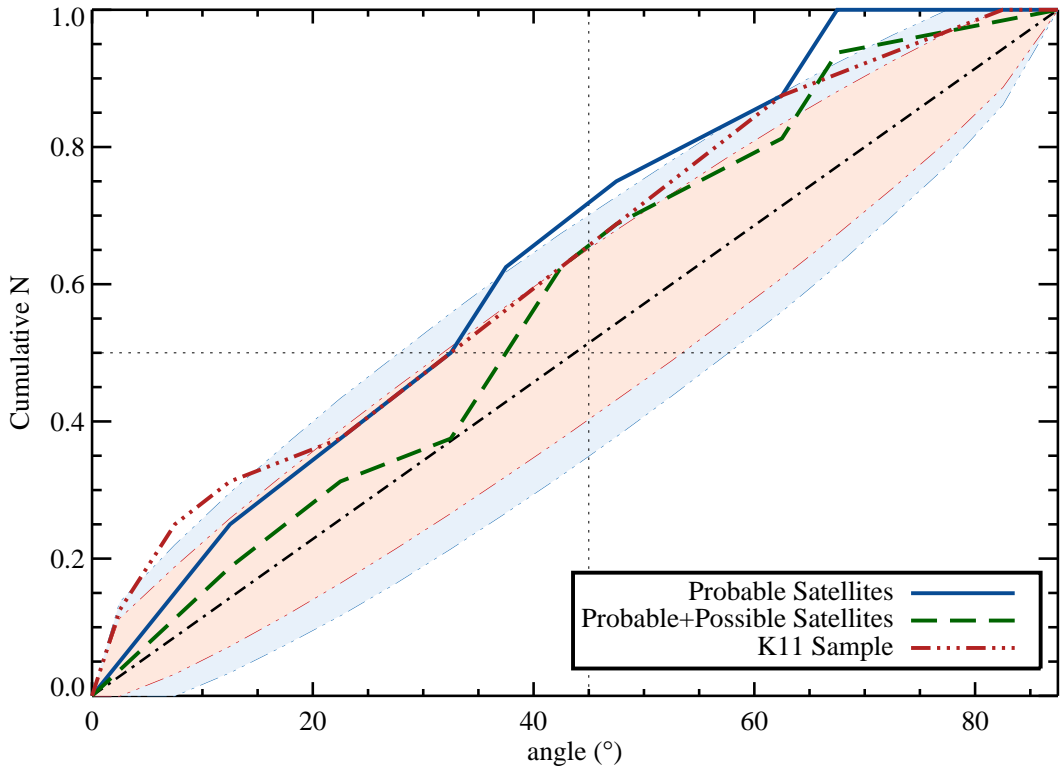}
\caption{Cumulative histogram showing angle between each satellite and
  the host's disk. The solid blue line indicates the 8 probable (``Y")
  satellite  galaxies  found in  this  study;  the  dashed green  line
  indicates the 8 probable plus  8 possible satellite galaxies (``Y" +
  ``M");  and the  dash-dotted  red line  indicates  the 16  satellite
  galaxies from K11. The blue and red shaded regions are the one sigma
  envelopes for  random distributions  drawn from samples  of 8  or 16
  satellites respectively.  The blue  and green samples  of satellites
  are  consistent with there  being no  angular dependence  around the
  disk of NGC  4258, shown as a straight  black dash-dotted line along
  the center of the shaded region.
 \label{angle}}
\end{figure*}

Figure  \ref{angle} displays  the cumulative  angular  distribution of
three subsets of  satellites. The blue solid line  traces the probable
satellites; the  dashed green line  traces the probable  plus possible
satellites;  and the  red  dash-dotted line  traces  the original  K11
satellites. Data  has been  folded from 360  degrees to 90  degrees to
allow  for  better  statistical  sampling.   An  angle  of  0  degrees
indicates a satellite galaxy is perfectly aligned with the disk of the
host (i.e. major axis alignment); an angle of 90 indicates a galaxy is
not at all aligned with the disk and might instead be aligned with the
minor axis  of the host.   In Figure \ref{angle}, a  black dash-dotted
line marks  the case  where there is  no angular dependence,  that is,
there  are just  as  many satellites  at  low angles  as high  angles.
Distributions  that  grow  faster  than  this  line  are  said  to  be
preferentially aligned  with the disk; distributions  that grow slower
than this line are not aligned with the disk.

Overplotted  on Figure  \ref{angle} are  the one  sigma  envelopes for
random distributions  drawn from samples  of eight and  16 satellites,
shown  in  blue  and  red  respectively. Our  probable  sample  has  8
satellites; the probable plus possible  sample and the K11 sample both
have  16 satellites.  All  three distributions  grow  faster than  the
random  distribution;  however both  the  probable  and probable  plus
possible  samples fall (marginally)  within the  envelopes of  what is
allowed by a random distribution.

For  completeness, we  have  used the  IDL  routine $ksone$  to run  a
one-sample Komolgorov-Smirnov (KS) test  on the data to quantitatively
assess if  the angular distributions of  probable satellites, possible
plus probable satellites,  and K11 satellites are drawn  from a random
distribution.   However,  we  caution  that  the KS  test  requires  a
relatively large  number of  data points to  properly reject  the null
hypothesis  (that   the  sample  is   drawn  from  a   random  uniform
distribution).

The  KS statistic ($D$)  specifies the  maximum deviation  between the
data and a supplied distribution;  $D$ varies between 0 and 1.  Larger
$D$  values  indicate that  the  data  and  supplied distribution  are
significantly different.  The significance level ($p$-value) of the KS
statistic  is also  considered; the  $p$-value is  the  probability of
drawing from a random distribution and obtaining results as extreme or
more extreme  than the  data.  $P$-values  vary from 0  to 1;  a large
$p$-value indicates  that it is  highly likely that one  will generate
samples like the data.

We make two  comparisons to a flat distribution,  where a satellite is
equally likely to be at any angular position: we consider our catalogs
of probable satellites, and probable plus possible satellites.  The KS
statistics for these are $D$=0.24 and 0.18 with significance levels of
$p$=0.66 and 0.63  respectively.  Since the $D$ numbers  are small and
$p$-values are  large, this  means with a  high confidence  level, our
samples of  probable and probable  plus possible satellites  are drawn
from a flat distribution.

Ever   since   Holmberg  initially   found   satellite  galaxies   are
preferentially   aligned  along   the   minor  axis   of  their   host
\citep{holmberg69},  there have  been a  steady stream  of conflicting
results   regarding   satellite   galaxy  alignments   \citep[see][and
  references   therein]{Bailin07}.   Besides  numerous   observational
disagreements, there is also no broad theoretical agreement on whether
satellites  should be  aligned at  all or  found on  randomly oriented
orbits   \citep{Zentner05,kroupa2005}.    There   are   two   standard
interpretations of satellites existing in a disk; either they recently
merged as  an infalling group,  or they tend  to fall along  cold dark
matter filaments  \citep{hartwick2000}. On the  other hand, a  lack of
any orientation could imply that  the host galaxy has not accreted any
new dwarfs  in recent  cosmic times. While  our findings  very loosely
support the latter hypothesis, we feel that we have too few satellites
to  draw definitive  conclusions.  There may  even  be an  alternative
interpretation  when taking satellite  colors into  consideration (see
Section \ref{satellite morphology}).

\subsection{Satellite Morphology}\label{satellite morphology}

\begin{figure*}
\centering
\plotone{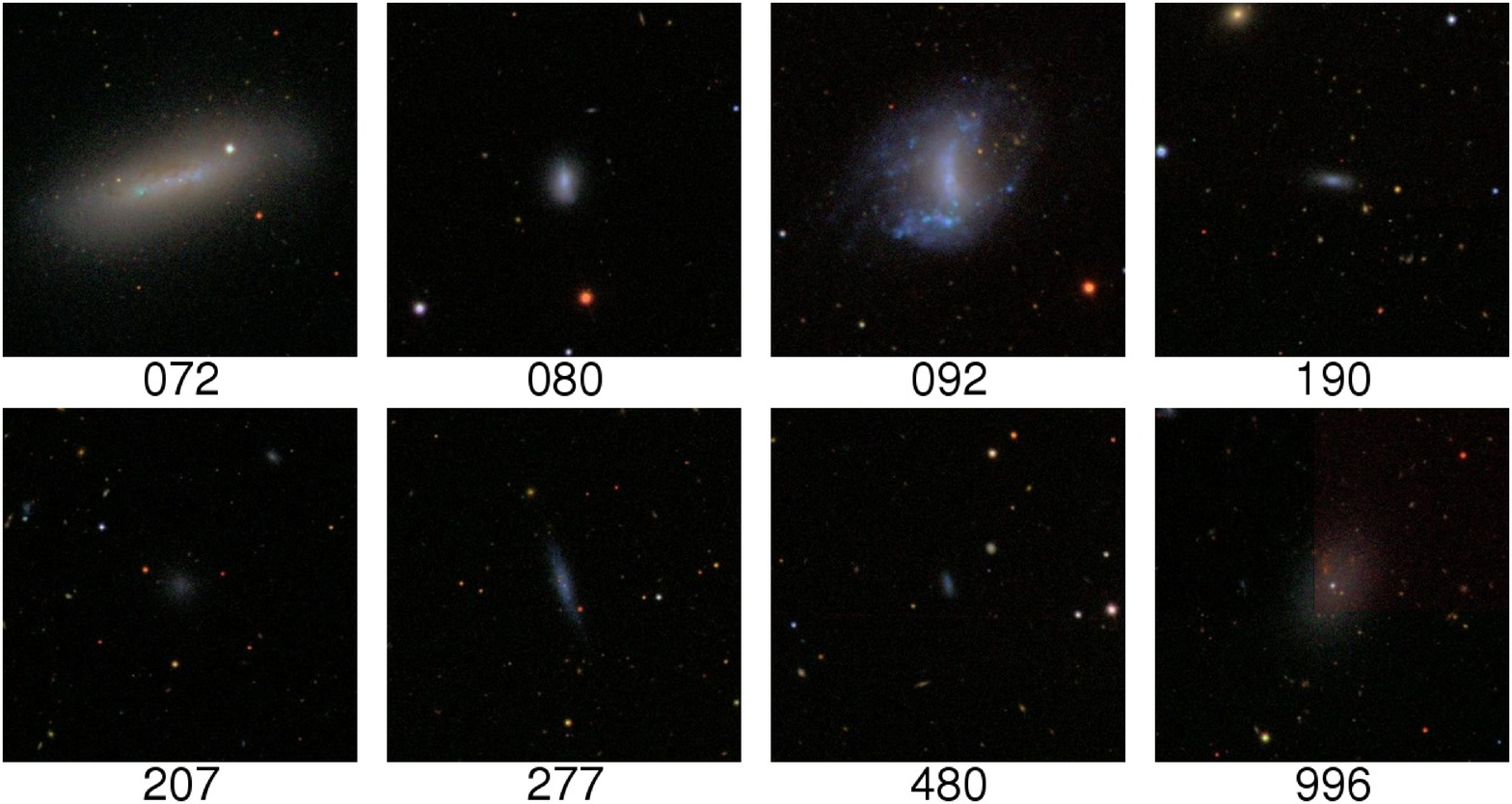}
\caption{SDSS  images of  probable (``Y")  satellites. The  last three
  digits of  each SDSS ID are  listed below the  respective image. The
  images are scaled to be 3.38 arcmin in diameter. }
\label{yes}
\end{figure*}

\begin{figure*}
\centering
\plotone{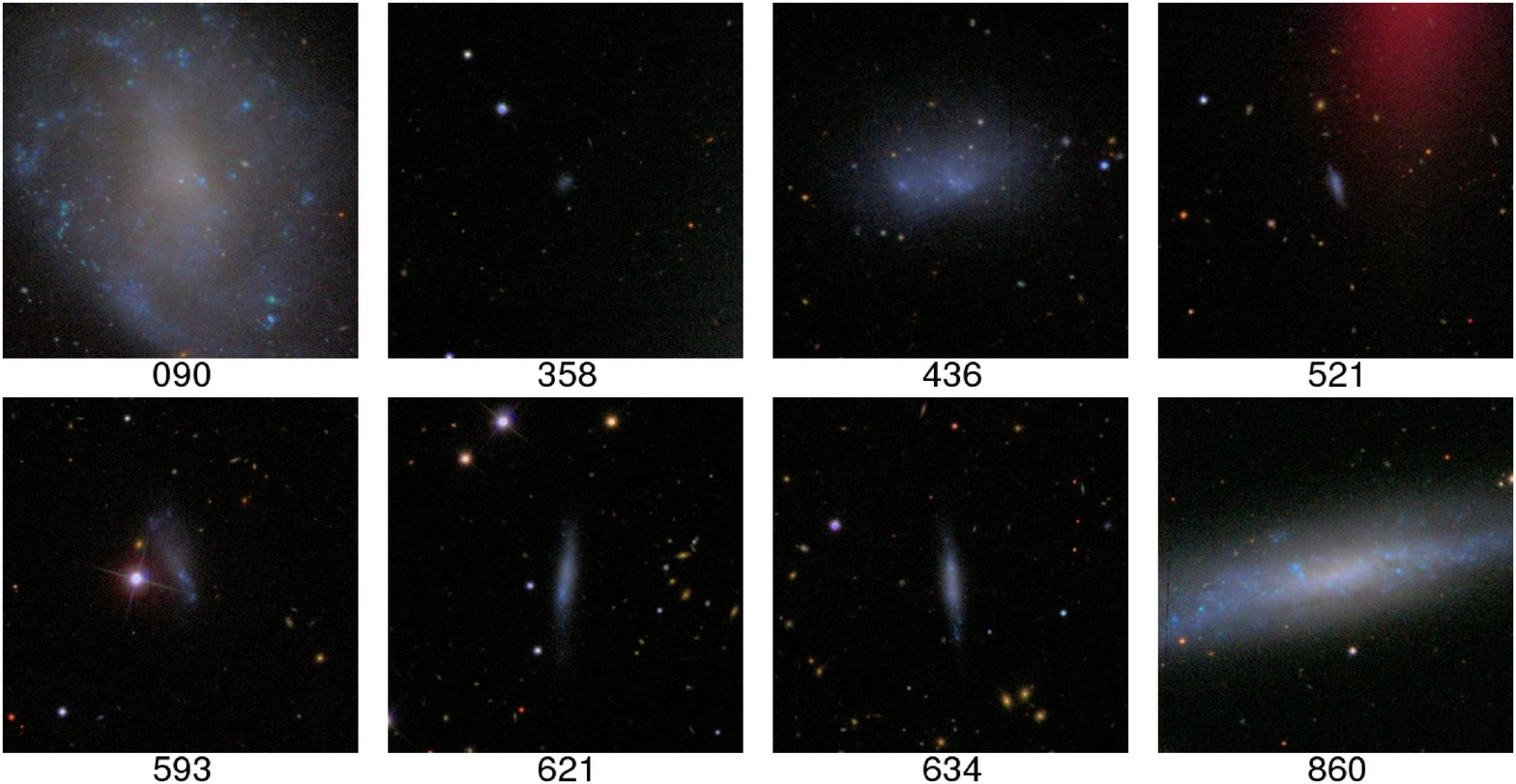}
\caption{SDSS  images of  possible (``M")  satellites. The  last three
  digits of  each SDSS ID are  listed below the  respective image. The
  images are scaled to be 3.38 arcmin in diameter. }
\label{maybe}
\end{figure*}

Figures  \ref{yes} and  \ref{maybe}  display the  SDSS  images of  our
probable and  possible satellites respectively. From  these images, is
immediately apparent  that the vast  majority of these  satellites are
blue irregulars.  We  plot further in Figure \ref{color}  a map of the
Sloan $u-r$ colors of the satellites as they appear on the sky. Colors
less  than  2.2 are  blue,  late-type galaxies.  All  but  two of  our
probable satellites have $u-r$ colors less than 1.7. The remaining two
have  colors less  than  2.15. Adding  to  that, all  of the  possible
satellite  galaxies  have  $u-r$  colors  less  than  1.6.  Given  our
technique  for  identifying  the  line  of  sight  velocity  of  these
galaxies,  this  is  not  surprising  (that is,  our  catalog  is  not
complete, as the method is  biased against red, dim satellites).  What
is surprising  is the sheer number  of blue satellites;  if this system
was like the  MW or M31, we would expect many  fewer blue star forming
satellites at small projected radii.

\begin{figure*}
\centering
\plotone{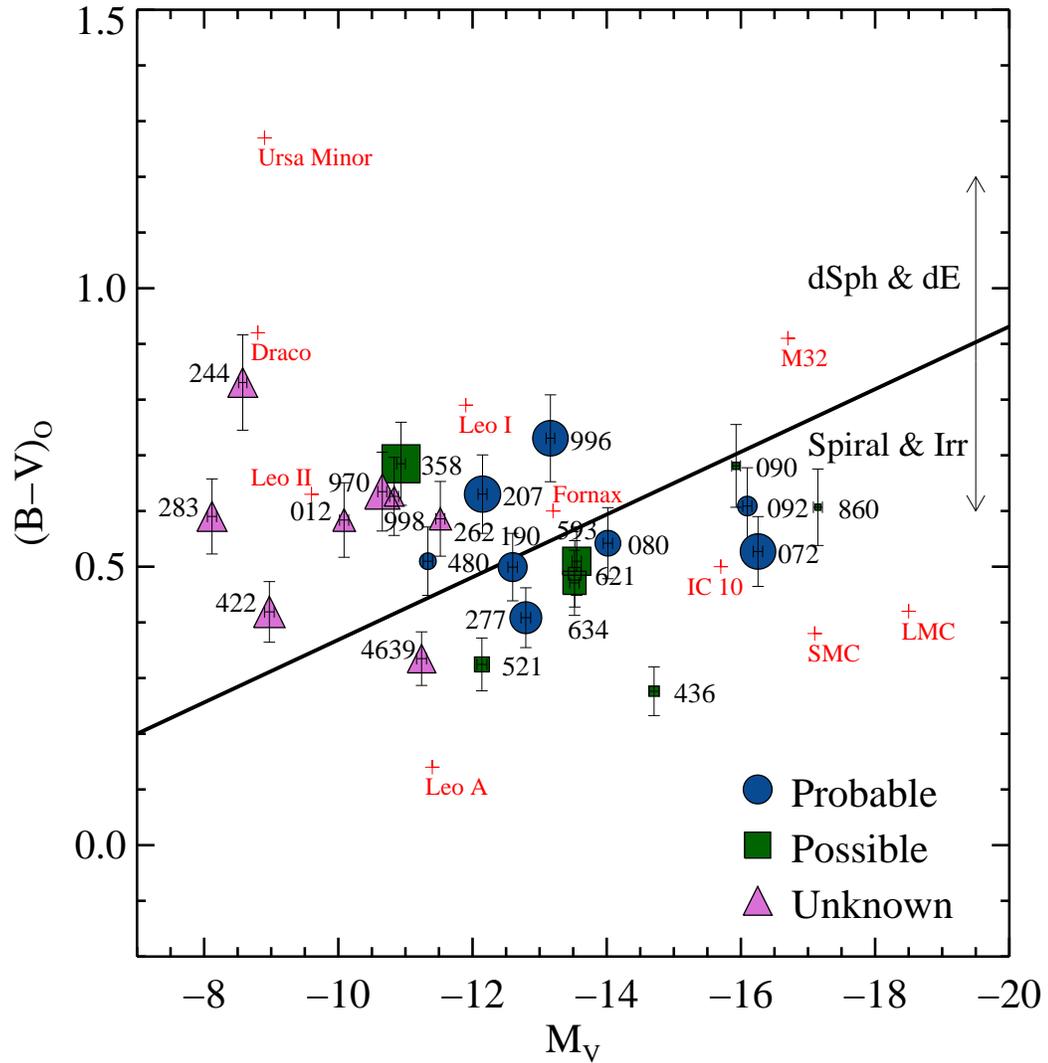}
\caption{Extinction  corrected  B-V  colors  of the  galaxies  plotted
  against absolute V-band magnitude.  Blue circles are probable (``Y")
  satellites, green  squares are possible (``M")  satellites, and pink
  triangles are galaxies that lack velocity information (``X"). Symbol
  size indicates projected radius; large symbols are satellites with a
  small projected  radius; small symbols  are satellites with  a large
  projected  radius.  Several  LG  dwarf  galaxies are  shown  as  red
  pluses. Galaxies below the line  are said to be blue; galaxies above
  the line are said to be red.}
\label{bvcolor}
\end{figure*}

To further stress this point, we replicate a color-magnitude plot from
\citet{mateo1998} shown in Figure \ref{bvcolor}. Color transformations
from  \citet{chonis2008} are  used  to move  from  g and  r  to B  and
V.  Galaxies  that  fall  below  the black  line  are  blue  late-type
galaxies;   galaxies  above   the  black   line  are   red  early-type
galaxies. Many of  our probable and possible galaxies  that have small
projected radii are found below the line.

It  is conventionally  thought that  galaxies experience  quenching as
they    fall    inward     toward    the    host    \citep[see,    for
  example,][]{geha2012}.  Since our  satellites are  blue star-formers
they most likely have not had enough time to be quenched. As mentioned
in the previous section, this might imply that they have recently been
accreted to the system.

Again, we are not making  an argument for completeness here because we
are  only sensitive  to the  brightest satellites  and  therefore miss
dimmer  dwarf spheroidals. However,  even in  an incomplete  sample we
find  a large  number of  blue,  late-type galaxies.  Our results  are
consistent  with the  photometric  work of  \cite{ludwig2012} for  NGC
7331.

\section{Discussion and Conclusion} \label{conclusion}
\subsection{Necessity of Spectroscopy}
As  we  have  demonstrated,  satellite membership  is  challenging  to
determine. It  cannot be done from photometry  alone, nor spectroscopy
alone,  but  instead  requires,  at  minimum,  a  combination  of  the
two. With  photometry, one  can make  a good first  guess as  to which
galaxies  might be  satellites  based on  resolvability  of stars  and
surface brightness  as was done  by K11. Spectroscopy can  help narrow
that sample, as  we have shown here. To  further refine the selection,
distance measurements  are needed.  Finally, to verify  a galaxy  as a
satellite  with absolute  certainty would  require orbital  radius and
proper motions. Without these, it is impossible to confidently declare
satellite membership.

While our sample of satellites does take velocities and distances into
consideration, we  caution that it might still  contain non-members in
the foreground  and background. Since  small number statistics  are in
play for  these sorts  of systems, it  only takes a  couple non-member
galaxies to  produce misleading conclusions  about the characteristics
of the sample  as a whole. This is why  we choose four categorizations
for our satellites: probable, possible, non-member, and unknown. We do
not refer to any of the satellites as confirmed. 

Despite our  cautionary language, we cannot emphasize  enough that our
sample  has  much better  constraints  than  a purely  photometrically
derived  ground-based satellite  catalog.  That  is, while  it  is not
complete and perhaps suffers  some minimal degree of contamination, it
is certainly the most reliable catalog of NGC 4258 satellites to date.

\subsection{Conclusion}
We present  a spectroscopic catalog  of 47 dwarf  galaxies surrounding
NGC 4258. Fifteen  of these targets did not  previously have published
redshifts.  A  histogram of  line  of  sight  velocities of  potential
satellite   galaxies  indicates  that   a  substantial   fraction  are
background  galaxies;   without  proper  motions   (or  realistically,
distances to the galaxies) there is no easy way to determine which are
bound to NGC 4258 and  which are background contaminants. Using an NFW
profile to  eliminate any  obvious interlopers, we  classify 8  of our
dwarf   galaxies   as   probable   satellites  and   8   as   possible
satellites.   Our   selection  criteria   are   based  upon   distance
measurements and  velocities. With this sample  of satellite galaxies,
we make four conclusions:

(1) The mass yielded when using the 8 probable satellite galaxies in a
mass   estimator   based   on   the  spherical   Jeans   equation   is
3.1$\pm$0.7$\times$10$^{12}$  M$_{\odot}$  out  to  a  radius  of  200
kpc. If we  instead integrate our NFW profile, we find  the mass to be
1.8$\times$10$^{12}$ M$_{\odot}$ out to 200 kpc.

(2) The orientation of the  probable and possible plus probable galaxy
subsets do not indicate a strong preferential alignment with the disk. 

(3) A  large number  of the probable  satellites are  blue irregulars,
which is atypical in comparison to the MW and M31 systems.

(4) Satellite membership is difficult to identify when only photometry
is  utilized. We  conclude  that velocity  and  distance measures  are
necessary to determine satellite membership with any certainty.

\begin{figure*}
\centering
\plotone{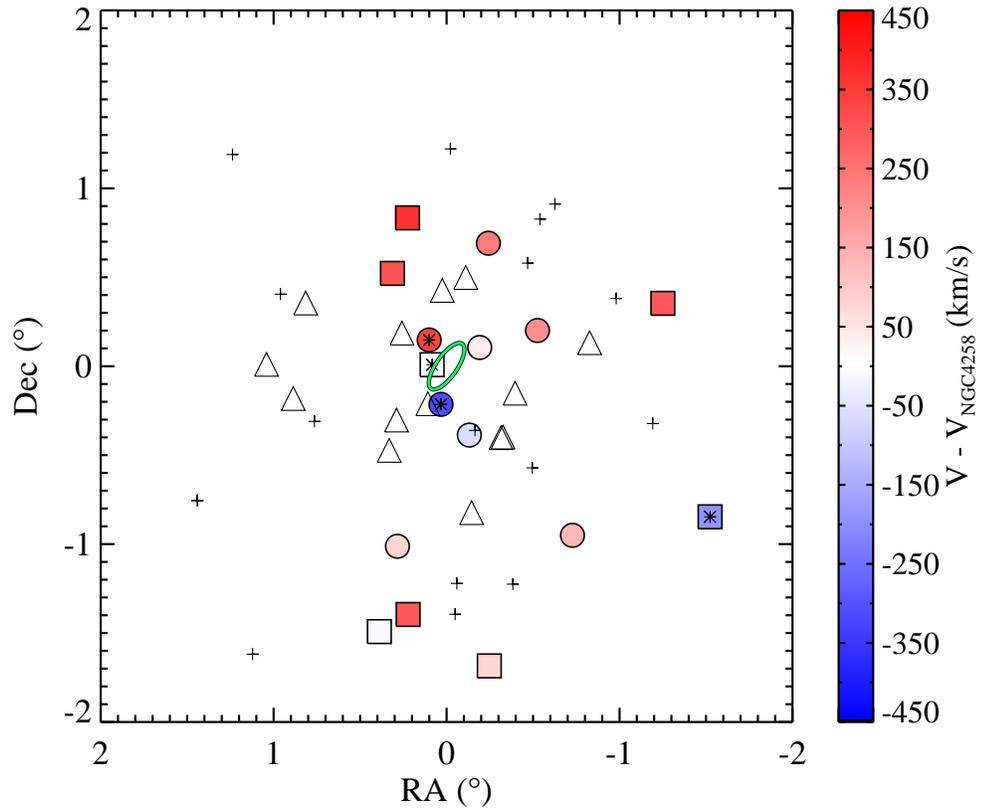}
\caption{Line of  sight velocity for  each satellite as it  appears in
  the sky. Red  indicates the satellite is redshifted  with respect to
  the host;  blue indicates the satellite is  blueshifted with respect
  to  the  host. The  green  ellipse  marks  the size,  location,  and
  inclination of NGC 4258. Circles are probable satellites (classified
  as  ``Y"); squares  are  possible satellites  (classified as  ``M");
  pluses are  background galaxies (classified as  ``N"); triangles are
  galaxies that  lack velocity  information (classified as  ``X"). The
  asterisks mark  galaxies with TRGB  distances; the two  circles with
  asterisks    are   the    galaxies   we    used   to    select   NFW
  parameters. \label{velocityfield}}
\end{figure*}

\begin{figure*}
\centering
\plotone{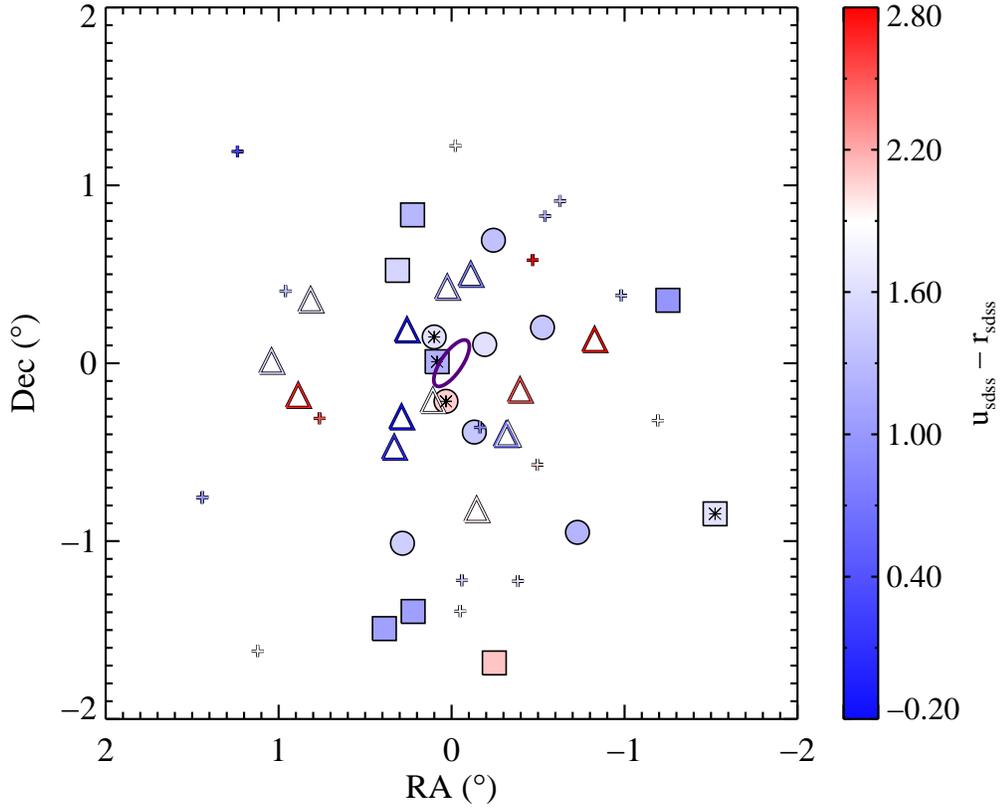}
\caption{SDSS  $u-r$ colors  for satellites  as seen  on the  sky. Red
  colored  symbols  represent redder  galaxies;  blue colored  symbols
  represent bluer  galaxies. Symbol shapes  are the same as  in Figure
  \ref{velocityfield}.  Most  of the  probable  (``Y") satellites  are
  bluer.   This    is   in   contrast   to    the   MW/M31   satellite
  system. \label{color}}
\end{figure*}

\clearpage

\begin{turnpage}
\begin{deluxetable}{ c c c c c c c c c c c}
\tabletypesize{\scriptsize}
\tablewidth{0pt}
\tablecaption{Summary of  galaxy properties from  our observations and
  the  literature.  All  velocities are  relative to  the heliocentric
  radial velocity. Column 3 is the  ID given by K11; entries of N/A in
  the APO velocity column are  galaxies that we observed but could not
  extract a radial velocity  for; entries of ``$>$10,000" are galaxies
  that we observed  that were high redshift; the  Status column is the
  member categorization  that we  describe in Section  3.1. ``Y"  is a
  probable  satellite; ``M"  is a  possible satellite;  ``N" is  not a
  satellite; ``X"  is a galaxy  lacking velocity information;  ``H" is
  the host galaxy. \label{data} }
\tablehead{\colhead{SDSS ID} & \colhead{Other ID} & \colhead{K11 ID} & \colhead{RA} &\colhead{Dec} & \colhead{APO v} & \colhead{SDSS v} &\colhead{Lit v} & \colhead{Dist} &\colhead{$r$} & \colhead{Status}\\
\colhead{(DR7)}& & &\colhead{(J2000)} & \colhead{(J2000)} & \colhead{(\kms)} & \colhead{(\kms)}& \colhead{(\kms)} & \colhead{(Mpc)} & \colhead{(mag)} 
}
\startdata
& NGC4258 &  & 184.74000 & 47.30400 & 451 &  & 456 & 7.60\tablenotemark{c} &
10.10 & H  \\
\hline
588017111295918190 &  SDSS J121551.55+473016.8  &  & 183.96482 & 47.50469 &  &
654 & 650 &  & 16.74 & Y  \\
588017109685633092 &  NGC 4288  &  & 185.15895 & 46.29180 & 527 & 522 & 528 &
8.05\tablenotemark{b} & 13.18 & Y  \\
588298663036846207 &  SDSS J121933.21+472705.2  & S8 & 184.88838 & 47.45147 &  &
786 & 788 & 7.05\tablenotemark{d} & 17.13 & Y  \\
588298663036715072 &  NGC 4248  & S4 & 184.45768 & 47.40920 & 482 &  & 492 &
7.40\tablenotemark{b} & 13.08 & Y  \\
588017110759243996 &  & S7 & 184.78782 & 47.08978 &  & 136 &  &
7.24\tablenotemark{b} & 16.09 & Y  \\
588298662499844277 &  SDSS J121811.04+465501.2  & S5 & 184.54613 & 46.91686 &
390 & 387 & 480 & 6.54\tablenotemark{b} & 16.59 & Y  \\
588017109685174480 &  &  & 183.66661 & 46.35330 & 583 &  &  &  & 17.99 & Y  \\
588017111832920080 &  2MASX J12173195+4759420  &  & 184.38344 & 47.99523 & 684 &
705 & 702 &  & 15.30 & Y  \\
\hline
588298664110653634 &  UGC 07392  & S10 & 185.07288 & 48.13766 & 819 & 805 & 805
&  & 15.82 & M  \\
588017111295590521 &  SDSS J121134.99+473927.1  &  & 182.89582 & 47.65755 & 749
&  & 755 &  & 17.30 & M  \\
588017605758222436 &  &  & 185.31344 & 45.81202 & 439 & 444 &  &
7.06\tablenotemark{b} & 14.72 & M  \\
588017627228930090 &  NGC 4242  &  & 184.37573 & 45.61930 &  & 528 & 514 &
7.90\tablenotemark{b} & 13.32 & M  \\
588017627765997621 &  UGC 07391  &  & 185.06768 & 45.90840 & 751 & 619 & 620 & 
& 15.81 & M  \\
588297863121272860 &  NGC 4144  &  & 182.49338 & 46.45740 & 261 & 263 & 273 &
7.24\tablenotemark{b} & 12.14 & M  \\
588298663036846358 &  &  & 184.86350 & 47.31255 &  &  &  & 7.05\tablenotemark{d}
& 18.32 & M  \\
588298663573782593 &  UGC 07401  & S13 & 185.20176 & 47.82592 & 759 & 757 & 770
&  & 15.78 & M  \\
\hline
588298663573717422 &  [KK98] 132  & S6 & 184.77705 & 47.73024 &  &  &  &  &
20.40 & X  \\
588017110759113395 &  & P5 & 184.15612 & 47.15199 &  &  &  &  & 22.38 & X  \\
588017110222504639 &  & S14 & 185.22911 & 46.83067 &  &  &  &  & 18.17 & X  \\
588298662499779244 &  & P4 & 184.25990 & 46.90557 &  &  &  &  & 20.62 & X  \\
588298663036912678 &  & S11 & 185.12007 & 47.49025 &  &  &  &  & 23.38 & X  \\
588298663037239998 &  &  & 186.27526 & 47.31420 & N/A &  &  &  & 18.44 & X  \\
588298662499779283 &  & P3 & 184.27664 & 46.90220 &  &  &  &  & 21.19 & X  \\
588298663573651859 &  & P1 & 184.57699 & 47.80259 &  &  &  &  & 23.92 & X  \\
588017110759309970 &  & S9 & 184.90003 & 47.09313 &  &  &  &  & 18.64 & X  \\
588017111296508012 &  & S16 & 185.94219 & 47.65887 &  &  &  &  & 19.20 & X  \\
588017110759637262 &  &  & 186.04768 & 47.12249 & N/A &  &  &  & 17.77 & X  \\
588297864195342828 &  & S1 & 183.52064 & 47.43611 &  &  &  &  & 18.98 & X  \\
588298662500041067 &  & S12 & 185.16674 & 47.00117 & N/A &  &  &  & 23.99 & X
 \\
588298661962974110 &  [KKH2011] P2  & P2 & 184.52705 & 46.48063 &  &  &  &  &
22.25 & X  \\
\hline
588017111295721701 &  &  & 183.29386 & 47.68492 & $>$10,000 &  &  &  & 18.36 & N
 \\
588017112369857019 &  &  & 184.70784 & 48.52569 & 13007 &  &  &  & 18.13 & N  \\
588017112370380911 &  UGCA 281  &  & 186.56543 & 48.49401 & 268 & 364 & 289 &
5.63\tablenotemark{b} & 15.28 & N  \\
588298662499844430 &  &  & 184.49752 & 46.94287 & 1149 &  &  &  & 17.98 & N  \\
588017606294700206 &  UGC 07301  &  & 184.17534 & 46.07877 & 709 &  & 698 &
21.50\tablenotemark{a} & 14.69 & N  \\
588017606294831288 &  &  & 184.65277 & 46.08295 & $>$10,000 &  &  &  & 18.33 & N
 \\
588017110222176377 &  MCG +08-22-086  &  & 184.00925 & 46.73285 & 1073 &  & 1065
&  & 15.37 & N  \\
588017627765866639 &  SDSS J121840.14+455434.9  &  & 184.66725 & 45.90971 &  &
1045 & 1062 &  & 16.33 & N  \\
588298663574110513 &  &  & 186.15665 & 47.70873 & $>$10,000 &  &  &  & 17.95 & N
 \\
588298664110260438 &  &  & 183.81594 & 48.21531 & $>$10,000 &  &  &  & 18.51 & N
 \\
588297863658340527 &  MCG +08-22-083  &  & 182.98213 & 46.98181 &  & 972 & 975 &
& 16.00 & N  \\
588298661963694280 &  &  & 186.86392 & 46.55062 & $>$10,000 &  &  &  & 18.18 & N
 \\
588298661963694346 &  &  & 186.86750 & 46.54675 & $>$10,000 &  &  &  & 18.65 & N
 \\
588017605758550042 &  &  & 186.39419 & 45.68523 & 717 & 712 &  &
14.40\tablenotemark{a} & 12.18 & N  \\
588298662500302850 &  NGC 4346  & S15 & 185.86649 & 46.99378 & 780 &  & 778 &
15.70\tablenotemark{a} & 11.41 & N  \\
588298664110325782 &  NGC 4218  & S2 & 183.94336 & 48.13084 & 685 & 796 & 738 &
21.10\tablenotemark{a} & 13.47 & N  \\
588298663573454909 &  NGC 4220  & S3 & 184.04879 & 47.88324 & 887 &  & 922 &
17.90\tablenotemark{a} & 11.58 & N  \\
\enddata
\tablenotetext{a}{NED}
\tablenotetext{b}{\citet{kara2013}}
\tablenotetext{c}{\citet{humphreys2013}}
\tablenotetext{d}{\citet{munshi2007}}
\end{deluxetable}
\end{turnpage}

\clearpage

\acknowledgments
\section{Acknowledgements}

The authors  would like to thank  L.~Macri and F.~Munshi for distance
measurements to  SDSS IDs 207 and  358. They would also  like to thank
M.~Mateo and M.~Valluri for many useful conversations and feedback, as
well as E.~Bell and C.~Slater for the  insight into distance measures
and  the reference  to the  \citet{kara2013} catalog  of  Local Volume
galaxies. The authors also thank  the anonymous referee for the useful
comments. SL acknowledges funding from the Michigan Society of Fellows
and thanks the ever cromulent S.~Garner.

This work was  based in part on observations  obtained with the Apache
Point Observatory 3.5-meter telescope,  which is owned and operated by
the  Astrophysical Research  Consortium.  We thank  the APO  observing
specialists for their help in executing these observations.

GANDALF was  developed by  the SAURON team  and is available  from the
SAURON     website    (www.strw.leidenuniv.nl/sauron).     See    also
\citet{Sarzi06} for details.

This research made use  of the NASA/IPAC Extragalactic Database (NED),
which  is  operated  by  the  Jet  Propulsion  Laboratory,  California
Institute of Technology, under  contract with the National Aeronautics
and Space Administration.

Funding  for  SDSS-III  has  been  provided by  the  Alfred  P.  Sloan
Foundation,  the  Participating  Institutions,  the  National  Science
Foundation, and the  U.S. Department of Energy Office  of Science. The
SDSS-III web site is http://www.sdss3.org/.

SDSS-III is  managed by the Astrophysical Research  Consortium for the
Participating Institutions of the SDSS-III Collaboration including the
University of  Arizona, the Brazilian  Participation Group, Brookhaven
National   Laboratory,  University   of  Cambridge,   Carnegie  Mellon
University, University of Florida, the French Participation Group, the
German  Participation  Group,  Harvard  University, the  Instituto  de
Astrofisica   de   Canarias,   the  Michigan   State/Notre   Dame/JINA
Participation  Group,  Johns  Hopkins  University,  Lawrence  Berkeley
National Laboratory, Max Planck Institute for Astrophysics, Max Planck
Institute for  Extraterrestrial Physics, New  Mexico State University,
New  York  University,   Ohio  State  University,  Pennsylvania  State
University,  University  of   Portsmouth,  Princeton  University,  the
Spanish Participation Group, University  of Tokyo, University of Utah,
Vanderbilt   University,  University   of   Virginia,  University   of
Washington, and Yale University.

\end{document}